\documentclass[longbibliography,a4paper, amsfonts, amssymb, amsmath, reprint, superscriptaddress, showkeys, twoside,floatfix,longbibliography, aps, prl]{revtex4-2}
\usepackage[english]{babel}
\usepackage[utf8]{inputenc}
\usepackage{caption}
\usepackage[colorinlistoftodos, color=green!40, prependcaption]{todonotes}
\usepackage{amsthm}
\usepackage{mathtools}
\usepackage{physics}
\usepackage{xcolor}
\usepackage{graphicx}
\usepackage[left=23mm,right=13mm,top=35mm,columnsep=15pt]{geometry} 
\usepackage{adjustbox}
\usepackage{placeins}
\usepackage[T1]{fontenc}
\usepackage{lipsum}
\usepackage{csquotes}
\usepackage{mhchem}
\usepackage{comment}
\newcommand{\si}{Supplemental Information}

\usepackage[pdftex, pdftitle={Article}, pdfauthor={Author}]{hyperref} %
\begin{document}

\title{The Importance of Nuclear Quantum Effects for NMR Crystallography}

\author{Edgar A Engel}
\email[Correspondence email address: ]{eae32@cam.ac.uk}
\affiliation{TCM Group, Cavendish Laboratory, University of Cambridge, J. J. Thomson Avenue, Cambridge CB3 0HE, United Kingdom}

\author{Venkat Kapil}
\affiliation{Laboratory of Computational Science and Modeling, Institut des Mat\'eriaux, \'Ecole Polytechnique F\'ed\'erale de Lausanne, 1015 Lausanne, Switzerland}

\author{Michele Ceriotti}
\affiliation{Laboratory of Computational Science and Modeling, Institut des Mat\'eriaux, \'Ecole Polytechnique F\'ed\'erale de Lausanne, 1015 Lausanne, Switzerland}

\date{\today}

\begin{abstract}
The resolving power of solid-state nuclear magnetic resonance (NMR) crystallography depends heavily on the accuracy of computational predictions of NMR chemical shieldings of candidate structures, which are usually taken to be local minima in the potential energy.
To test the limits of this approximation, we systematically study the importance of finite-temperature and quantum nuclear fluctuations for $^1$H, $^{13}$C, and $^{15}$N shieldings in polymorphs of three paradigmatic molecular crystals -- benzene, glycine, and succinic acid.
The effect of quantum fluctuations is comparable to the typical errors of shielding predictions for static nuclei with respect to experiments, and their inclusion to improve the agreement with measurements, translating to more reliable assignment of the NMR spectra to the correct candidate structure.
The use of integrated machine-learning models, trained on first-principles energies and shieldings, renders rigorous sampling of nuclear fluctuations affordable, setting a new standard for the calculations underlying NMR structure determinations.

\end{abstract}


\maketitle

\setcounter{topnumber}{2}
\setcounter{bottomnumber}{2}
\setcounter{totalnumber}{8}
\renewcommand{\topfraction}{1.0}
\renewcommand{\bottomfraction}{0.85}
\renewcommand{\textfraction}{0.1}
\renewcommand{\floatpagefraction}{0.8}

\begin{figure}[ht]
    \centering
    \includegraphics[width=0.3\textwidth]{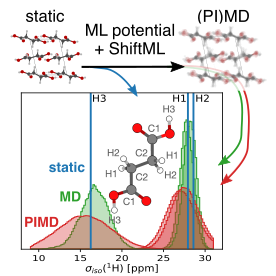}
    \captionsetup{labelformat=empty}
\end{figure}

Solid-state nuclear magnetic resonance (NMR) is a powerful technique for studying the microscopic structure and dynamics of amorphous and powdered samples~\cite{reif_2021}
and can resolve the positions of light elements such as hydrogen~\cite{florian2013}, which are difficult to observe with X-ray diffraction.
Chemical-shift based NMR crystallography exploits the sensitive dependence of the shielding of nuclear spins from external magnetic fields on the electron density surrounding the nuclei,  which makes it an accurate probe of their local atomic environments. 

The interpretation of solid-state NMR experiments often relies on comparisons with first-principles predictions of chemical shieldings, that are usually performed for the nuclear positions that correspond to a local minimum in the potential energy surface (``static'' calculations).
For gas-phase molecules, extensive studies highlight the impact of the quantum-mechanical and thermal motion of the constituent nuclei on such comparisons~\cite{gee2000, boehm2000, schulte2001, ruden2003, dracinsky2009, zhou-wang20jcp}, and indicate that they are comparable to the discrepancy between the most widespread electronic structure approaches based on density-functional-theory (DFT)~\cite{hohenberg1964, kohn1965, payne1992} gauge-including projector-augmented waves (GIPAW) schemes~\cite{thon+09jcp, bonhomme_2012, pickard_2001_gipaw, yates_2007_gipaw}, and more sophisticated and demanding quantum-chemistry techniques\cite{helgaker_2007, hartman_2017, hartman_2018, dittmer_2020}.
In the condensed phase, it has been shown that the thermal motion of nuclei is similarly important to compare with NMR experiments, and allows assessing the uncertainty in nuclear positions from NMR crystal structure determinations based on the discrepancies between the predicted and the measured chemical shifts~\cite{hofstetter_2017}.
Nonetheless, due to the high cost of NMR shielding calculations in the condensed phase, most studies thus far have either neglected nuclear quantum effects (NQE) altogether, or treated them subject to approximations to the description of nuclear motion and the dependence of shieldings on the displacements of nuclei from their equilibrium positions~\cite{rossano2005, dumez2009, schmidt2005, lee2007, robinson2010, gortari2010, hass+12jacs, ceri+13pnas, dracinsky2013, dracinsky2014, dracinsky2012, monserrat2014}.
These approximations are often used to compute approximate free energies for nearly harmonic systems, but have been found to be inadequate for general condensed-matter systems, such as molecular crystals, which exhibit strongly anharmonic behaviour due to molecular flexibility~\cite{reil-tkat14prl, rossi2016, ko+18prm, kapil2019, raim+19prm}. As we shall see, they are also not suitable for describing finite-temperature and quantum effects on NMR observables, and more accurate calculations based on the sampling of the quantum distribution using path integral molecular dynamics (PIMD)~\cite{mark-ceri18nrc} are needed. 
\begin{figure*}[bthp]
    \centering
    \includegraphics[width=\textwidth]{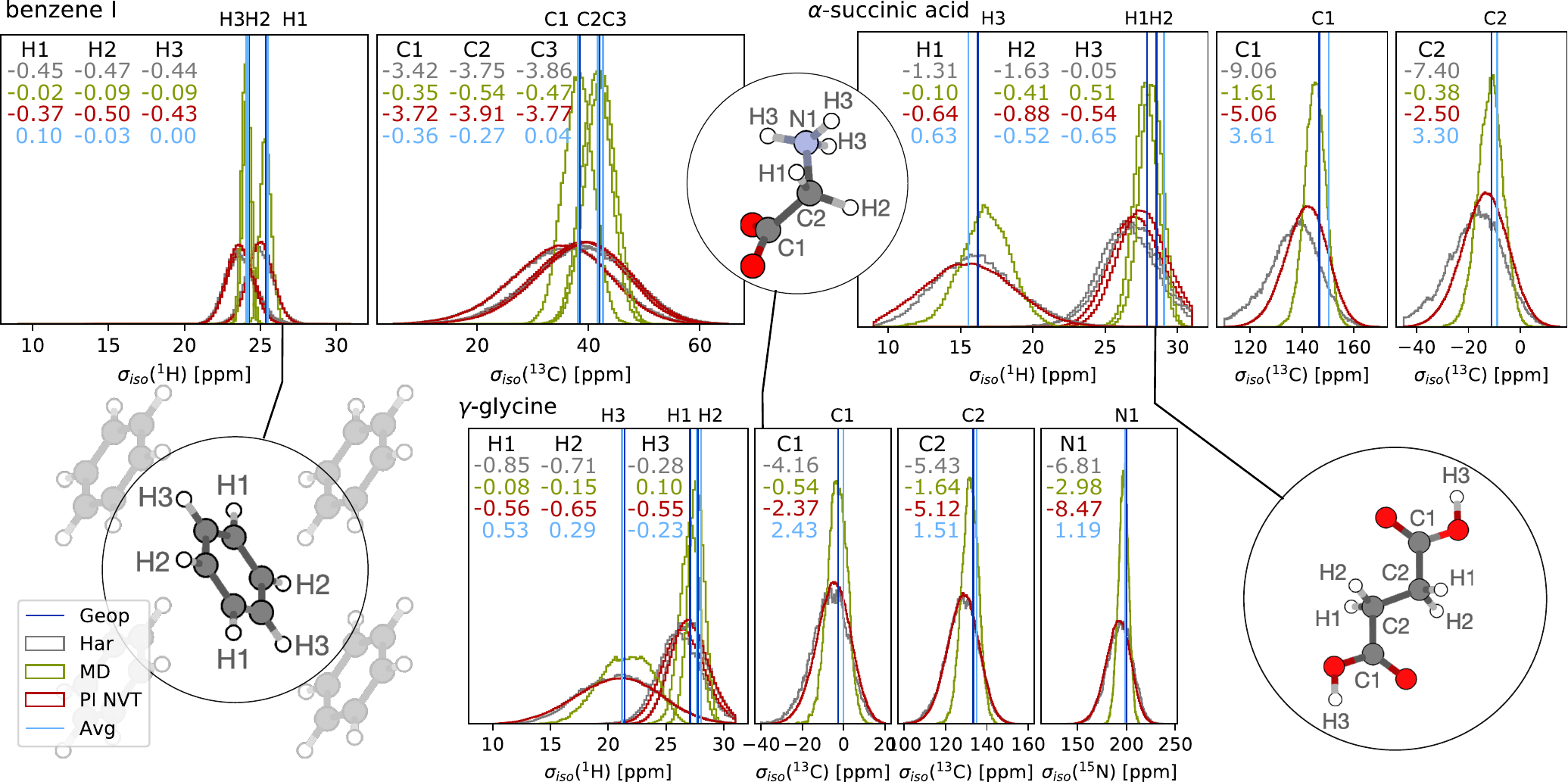}
    \caption{Distributions of the computed $^1$H, $^{13}$C, and $^{15}$N chemical shieldings, computed forr benzene form I, $\alpha$-succinic acid and $\gamma$-glycine.
Distributions are shown for the static (geometry optimised) structures (Geop), quantum Harmonic approximation (Har) MD with classical nuclei (MD), PIMD (PI NVT) and PIMD average structure (Avg). 
The molecular sites corresponding to each atom label are indicated in the depictions of the three compounds. The change in $\sigma_\text{iso}$ relative to the static structure also shown, in ppm, color-coded according to the sampling method. }
    \label{fig:nqe}
\end{figure*}

In this work we consider the low-energy crystalline polymorphs of three paradigmatic compounds: forms I and II of benzene~\cite{katrusiak2010, budzianowski2006}, $\alpha$- and $\beta$-succinic acid~\cite{dodd1998, leviel1981}, and $\alpha$-, $\beta$-, and $\gamma$-glycine~\cite{dawson2005}.
Even though these are model compounds that contain too few independent NMR signals to allow for conclusive structural determination, they are representative of molecular crystals dominated by van-der-Waals interactions, strong hydrogen bonds, and ionic interactions, and are well-suited to provide a demonstration of an integrated approach that combines sampling of the quantum distribution using machine learning models to compute both the potential the chemical shieldings.
We perform path integral molecular dynamics (PIMD) simulations in both the constant-volume ($NVT$) and constant-pressure, ($NpT$) ensembles, as implemented in i-PI\cite{kapil_2018_ipi}, allowing us to rigorously account for anharmonic quantum nuclear motion and the fluctuations and thermal expansion of the simulation cell.
For the sake of comparison, we also calculate the NMR signals for the ``static'' geometry, for a harmonic approximation of the distribution, for finite-temperature trajectories performed with classical MD, as well as for the ``mean configurations'' obtained by averaging the atomic coordinates over the respective PIMD trajectories.

We keep the cost of sampling affordable by using a machine-learning potential (MLP),\cite{behl-parr07prl,bart+10prl} based on the Behler-Parrinello framework~\cite{behl-parr07prl}. 
In the past decade, MLPs have simplified greatly the task of assessing finite-temperature thermodynamics of materials~\cite{chen+19pnas, jinn+19prl, deri+20nc, niu+20nc, imbalzano2021gaas}.
Here we use a MLP thas has been proven to accurately reproduce the reference first-principles configurational energies, and free energy differences between the above polymorphs~\cite{kapil2021}, trained on accurate DFT calculations using the hybrid PBE0 exchange-correlation functional~\cite{perdew1996pbe0, adamo1999pbe0} and a many-body dispersion (MBD) dispersion correction~\cite{tkatchenko2012mbd, ambrosetti2014mbd} (PBE0-MBD). More details on the construction of the training dataset and the MLPs may be found in Ref.~\cite{kapil2021}, and the associated data record\cite{matcloud21b}. 
We combine the MLP with machine learning models for the chemical shieldings, following a protocol similar to that described in Refs.~\cite{paruzzo_2018_shiftml, engel2019bnmr}, and training on GIPAW-DFT $^1$H, $^{13}$C and $^{15}$N chemical shieldings computed for configurations extracted from preliminary PIMD calculations.  
Separate models are built for each compound, which allows reaching very high accuracy, with average root-mean-square errors (RMSE) in $^{1}$H, $^{13}$C, and $^{15}$N predictions of 0.15\,ppm, 0.9\,ppm, and 1.3\,ppm, respectively. These errors are less than half of the literature estimates of the typical error of this flavour of GIPAW-DFT calculations for static, geometry-optimised configurations with respect to experiment~\cite{salager_2010_nmr, hartman_2016, dracinsky_2019}, and roughly half of the ShiftML model~\cite{paruzzo_2018_shiftml, engel2019bnmr}, which can make predictions across different compounds but is limited to structures that correspond to minima in the potential energy surface. The model also includes uncertainty quantification through a calibrated committee model~\cite{musil_2019_uncertainty}. 

\begin{figure*}[bthp]
    \centering
    \begin{minipage}{0.6\linewidth}
        \includegraphics[width=1.0\linewidth]{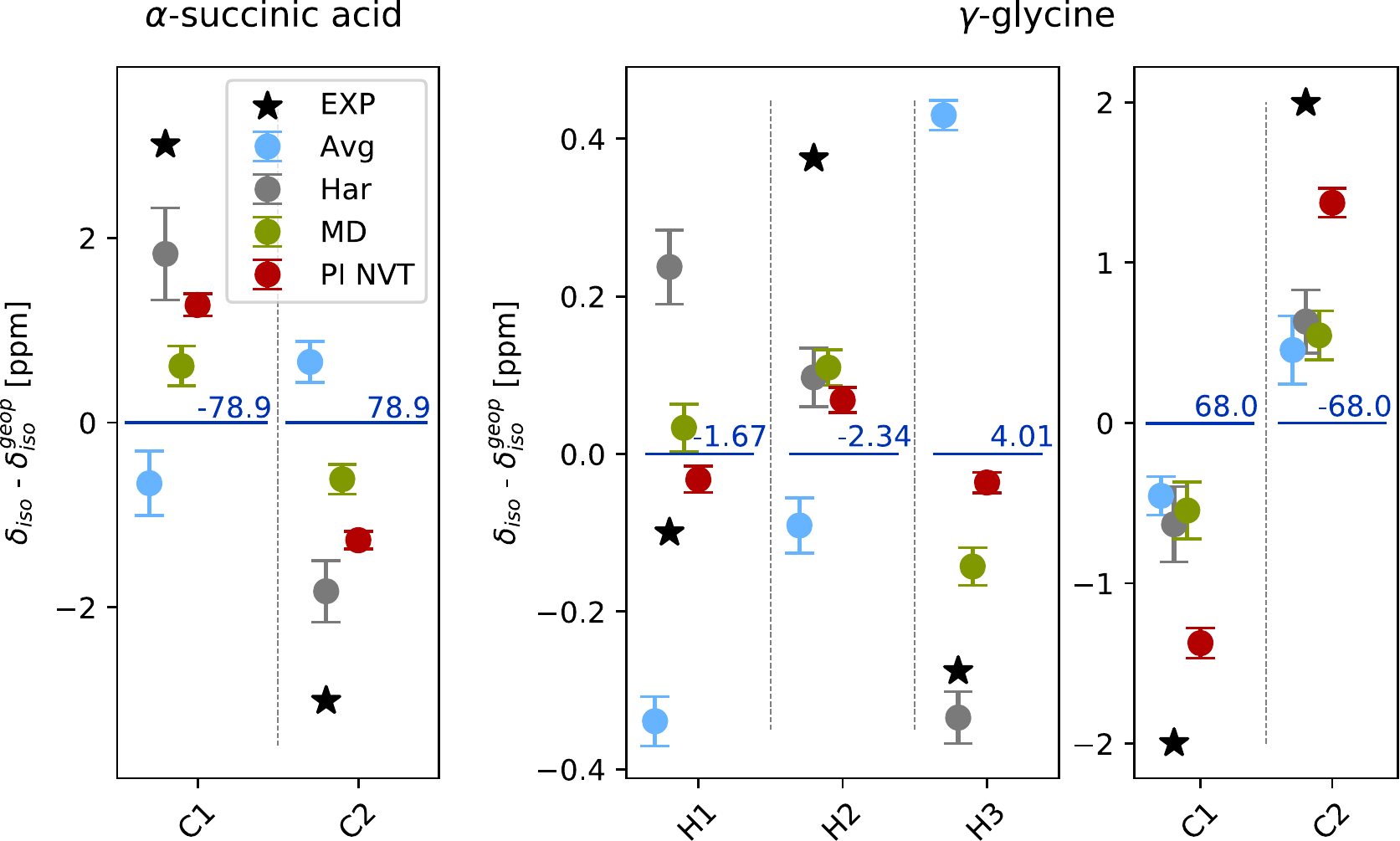}
    \end{minipage}\quad\quad
    \begin{minipage}{0.25\linewidth}
        \caption{Corrections to the chemical shifts with respect to the static structure of $\alpha$-succinic acid and $\gamma$-glycine, for the different sampling methods, color-coded as in Fig.~\ref{fig:nqe}, and internally-calibrated according to Eq.~\eqref{eq:shift-iso}. Error bars reflect ML uncertainties as well as, where applicable, statistical errors due to sampling. 
        Experimental values~\cite{jagannathan1987succinic, yamada2008glycine, folliet2013glycine, cerreia2018glycine, paruzzo2019glycine, castiglia2019glycine} are indicated as black stars.
        \label{fig:nqe_exp}}
    \end{minipage}
\end{figure*}
Readers interested in reconstructing or employing the shielding models are referred to the Materials Cloud archive~\cite{repository2}, which includes the (i) full GIPAW-DFT reference data, (ii) lighter, sparsified equivalents of the models themselves, and (iii) example python scripts for model generation and evaluation, as well as (iv) the MLPs and sample input for reproducing the statistical sampling simulations.

A rigorous evaluation of the observable value of the shielding involves averaging over the distribution of thermal (and quantum) fluctuations of the nuclei. 
In Figure~\ref{fig:nqe} we show the distributions of chemical shieldings obtained from classical MD, PIMD, as well as for a distribution of configurations based on the harmonic Hessian computed on the local minimum energy geometry. 
The most prominent effect is a broadening of the shielding distributions, associated with thermal and/or zero-point nuclear flucturations.  
Crucially, however, the \emph{mean} of the distributions shifts as well, by up to 1\,ppm for \ce{^1H} and 2\,ppm for \ce{^{13}C} and \ce{^{15}N}. 
No approximate technique quantitatively captures the change in $\sigma_\text{iso}$ associated with NQEs across the three systems. In particular, neither classical MD nor a Gaussian approximation of the fluctuations suffice to reproduce consistently PIMD results to within 1 ppm.
The large errors that are often associated with the use of the PIMD-averaged molecular geometry indicates that the dependence of $\sigma$ on fluctuations is non-linear.

\begin{figure*}[btp]
    \centering
    \includegraphics[width=1.0\textwidth]{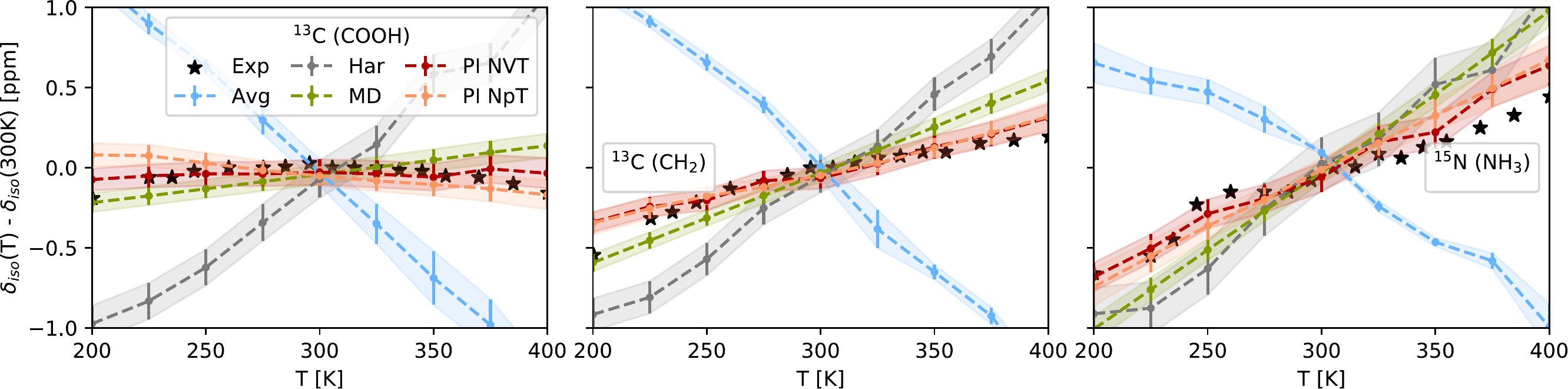}
    \caption{Temperature dependence of the experimentally distinct $^{13}$C and $^{15}$N $\delta_\text{iso}$ for $\gamma$-glycine, as obtained with different sampling methods, color-coded as in Fig.~\ref{fig:nqe} ($NpT$ PIMD is indicated as orange lines). Experiments from Ref.~\citenum{taylor2008} are shown as black stars.
    \label{fig:tempdep}}
\end{figure*}

Comparison with experiment requires translating the predicted chemical shieldings $\sigma$ into chemical shifts, $\delta_{\textrm{iso}}$, i.e. the change in the resonance frequencies of nuclei relative to a given reference, $\sigma_{\textrm{ref}}$.
\begin{equation}
\delta_{\textrm{iso}} = \sigma_{\textrm{ref}} - \operatorname{Tr}(\sigma) \label{eq:shift-iso}
\end{equation}
\emph{For each structure} and nucleus type we take $\sigma_{\textrm{ref}}$ to be the weighted mean of the shielding values. When comparing the values of $\delta_{\textrm{iso}}$ between different polymorphs, this choice is equivalent to the minimization of the discrepancy by a rigid relative translations of the peaks, which is common practice in solid-state NMR~\cite{salager_2010_nmr, hartman_2016}.   
The application of Eq.~\eqref{eq:shift-iso} to  compare the shieldings of different polymorphs, or to assess predictions against experiments, absorbs any rigid, global changes of the spectrum. Still, the degree to which any shielding is affected by fluctuations depends on its local environment. 
Fig.~\ref{fig:nqe_exp} shows a visual representation of the shifts for the polymorphs of succinic acid and glycine for which experimental data is available~\cite{jagannathan1987succinic, yamada2008glycine, folliet2013glycine, cerreia2018glycine, paruzzo2019glycine, castiglia2019glycine} (a full table is provided in the \si). 
Even after internal calibration, NQEs lead to changes of the chemical shifts by up to 0.2\,ppm for $^1$H and 2\,ppm for $^{13}$C, in every case reducing the discrepancy with measurements.
This is particularly apparent for $^{13}$C shifts, where incorporating NQEs reduces the discrepancy with experiments by up to 1\,ppm. This gives an indication of the potential improvements that can be expected by implementing a full PIMD description of the system, which may be made affordable by the development of more generally applicable integrated ML models, capable of simultaneously predicting both energetics and chemical shieldings for a broad class of materials. 
By considering the PIMD results as the ground truth, which is correct when assessing the error associated with the sampling of an approximate distribution, or with the reduction of sampling to a single structure, neither classical MD nor an harmonic approximation provide a systematic improvement relative to the static, geometry-optimized structures. This is similar to what was observed for the estimation of the free energy of polymorphs~\cite{kapi+19jctc}, suggesting that a rigorous description of nuclear motion is indispensable for moving beyond the static lattice approximation. 

\begin{figure}[hbtp]
    \centering
\includegraphics[width=0.9\linewidth]{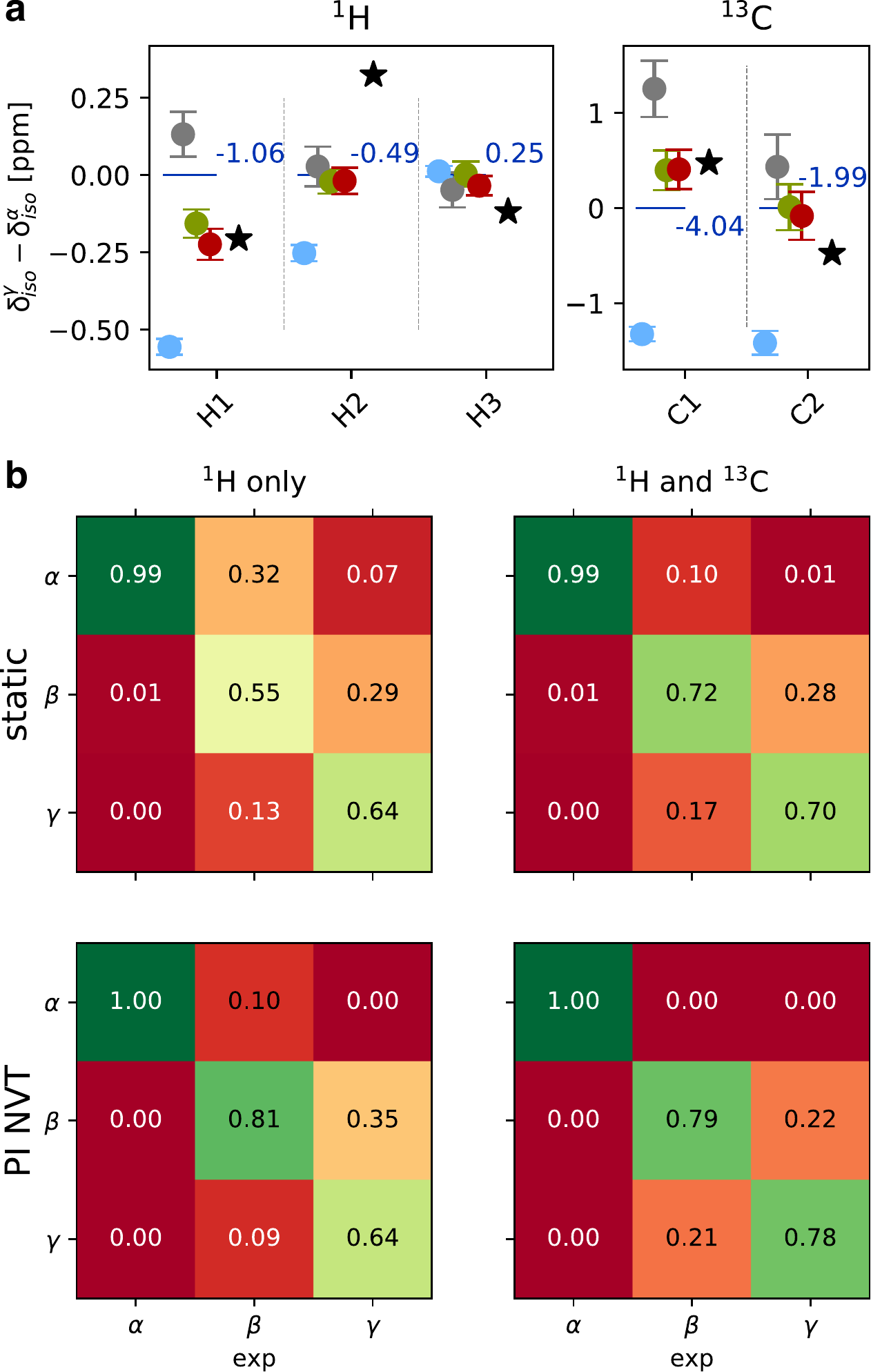}
\caption{a) Differences in chemical shifts between $\gamma$- and $\alpha$-glycine (figures for other compounds are in the \si). The results for static structures are used as reference, and the color coding of the other sampling methods follows that in Fig.~\ref{fig:nqe}.
Experimental observations are indicated by black stars. 
b) Bayesian probabilities of matching the crystal structures of $\alpha$, $\beta$, and $\gamma$-glycine to the respective experimentally measured $^1$H (left) and $^1$H and $^{13}$C (right) chemical shifts~\cite{yamada2008glycine, folliet2013glycine, cerreia2018glycine, paruzzo2019glycine, castiglia2019glycine}, as determined using the Bayesian framework of Ref.~\cite{engel2019bnmr}. 
The upper panels show the assignment probabilities using ML predictions for geometry-optimised structures, while  lower panels are based on shift predictions from PI simulations in the $NVT$ ensemble at 300\,K.
}    \label{fig:glycine-nmrx}
\end{figure}

The need to include nuclear fluctuations in the theoretical treatment of solid-state NMR is underscored by the experimentally-observed temperature dependence of chemical shifts.
In Fig.~\ref{fig:tempdep} we show results for the temperature-dependence computed for $\gamma$-glycine between 200 and 400~K. %
The increase in the $^{13}$C and $^{15}$N shifts of the methylene and amine groups, up to 0.7 and 1.2\,ppm respectively, may be understood in terms of the thermally-driven delocalisation of electron density, which leads to a reduction in chemical shielding.
Less intuitively the $^{13}$C shift associated with the carboxyl group decreases slightly above room temperature.
The $^1$H shifts exhibit relatively weak temperature dependence between 200 and 400\,K, with an overall increase of less than 0.05 ppm for the methylene $^1$H shift,  and a decrease of about 0.1 ppm for the amine $^1$H shift (see \si{}).
Among the different sampling methods, only PIMD achieves a quantitative agreement with the experimental observations,  and even the slight downward trend for the carboxyl $^{13}$C shift is matched quantitatively. The harmonic, and mean structure approximations lead to particularly large discrepancies.  
We also compare $NVT$ simulations performed using the experimental cell parameters and $NpT$ simulations. 
The mean volume obtained in constant-pressure simulations agrees with experiment to within 3\%, and accounts for the effects of fluctuations of the simulation cell and thermal expansion. 
This small difference is not reflected in significant changes in the predicted chemical shieldings for $\gamma$-glycine, indicating that the thermal expansion of the cell by around 1-2\% in volume does not significantly affect $\sigma$.
Even though the compounds we study here are usually considered too simple to be amenable to NMR crystallography, there is sufficient experimental data available for the three polymorphs of glycine to attempt structure determination, based on the distinct $^1$H and $^{13}$C shifts~\cite{yamada2008glycine, folliet2013glycine, cerreia2018glycine, paruzzo2019glycine, castiglia2019glycine}, so as to give an indication of the relevance of the NQEs-induced modulation of the chemical shieldings on NMR crystallography.
First, we observe that even though there is a degree of cancellation between the changes in $\delta_\text{iso}$ for pairs of polymorphs, the residual effect is still significant, up to 0.2 ppm for $^1$H and 0.5 ppm for $^{13}$C (Fig.~\ref{fig:glycine-nmrx}a), which is comparable to the typical discrepancy between static GIPAW-DFT calculations and experiment~\cite{salager_2010_nmr, hartman_2016, dracinsky_2019}. We also note that, among the approximate methods, only classical MD predicts the correct trend, although it usually underestimates the effect.

We then investigate whether these differences have a noticeable effect on the resolving power of NMR crystallography, using the probabilistic framework introduced in Ref.~\citenum{engel2019bnmr}.
In a nutshell, based on the discrepancy between computed and experimental shifts, and taking into account the uncertainty in the computed shifts, one can estimate the probability 
$p(\mathbf{y}|M)$ that the experimental shifts $\mathbf{y}$ are compatible with polymorph $M$ -- which can be interpreted as the level of confidence in the assignment. 
We assume the experimental shifts to be correctly assigned to the different NMR sites, and take uncertainties in shifts to arise from a combination of errors in the GIPAW-DFT reference and the uncertainty in their reproduction by the ML model. We therefore add in quadrature the uncertainties in the averaged ML predictions (that are obtained for each nucleus from a combination of sampling and ML errors) to a baseline value accounting for the errors in the GIPAW-DFT reference (a single value for all $^1$H and one for the $^{13}$C shifts, determined by likelihood maximization~\cite{engel2019bnmr}).
Fig.~\ref{fig:glycine-nmrx}b shows that (even though we are only considering three possible candidates) using the static-structure $^1$H shieldings only allows for a conclusive structural determination of $\alpha$-glycine, while the assignment of $\beta$ and $\gamma$ glycine is ambiguous. 
Combining $^1$H and $^{13}$C data improves the reliability of the assignment. The self-consistently optimized baseline uncertainties are found to be 0.29 and 2.0\,ppm for $^1$H and $^{13}$C, respectively, which is consistent with the typical discrepancies observed in the literature between experiments and static-structure GIPAW calculations~\cite{salager_2010_nmr, hartman_2016, dracinsky_2019}.

Accounting for NQEs leads to clearer assignment of the $\beta$ and $\gamma$ phases, both using only $^1$H shifts and using the combination of $^1$H and $^{13}$C values. In the latter case, the correct assignment probabilities increase to almost 80\%. 
Furthermore, the optimized baseline uncertainties are reduced to 0.24 and 0.57\,ppm for hydrogen and carbon, indicating that the Bayesian analysis benefits from the improvement of prediction accuracy afforded by PIMD averaging.
A more systematic study including more complex structures would be needed to confirm the generality of our findings, but the substantial reduction in the uncertainty associated with $^{13}$C shielding is consistent with the observed improved agreement with experimentally-measured shifts, and with the stronger sensitivity of carbon shieldings to temperature-induced fluctuations. 
In summary, we have demonstrated that computing finite-temperature, ensemble-average NMR chemical shifts for condensed-matter systems with first-principles accuracy is no longer the Herculean task that it used to be.
We present three paradigmatic examples demonstrating the use of an integrated ML model, which predicts both interatomic potentials and chemical shieldings, to assess the impact of thermal and quantum fluctuations of the nuclei on the accuracy of NMR shift predictions and the resolving power of NMR crystallography. 
We find that nuclear quantum effects change chemical shieldings by an amount that is comparable with the typical errors observed between DFT and experiments, and that significant corrections persist even when considering internally-calibrated shifts, and the comparison between polymorphs of the same compound. 
Comparison with experiments shows that NQEs are needed to achieve  quantitative accuracy in predicting the temperature dependence of shifts.
For the case of $\alpha$, $\beta$ and $\gamma$ glycine, we show that rigorous finite-temperature, ensemble-average chemical shifts lead to a reduction in errors with respect to experiment, and render otherwise ambiguous NMR crystal structure determinations conclusive and reliable.

A ML-powered simulation scheme with explicit sampling of quantum and thermal fluctuations offers several practical advantages. The fully-anharmonic configurational sampling is robust to molecular flexibility and dynamic disorder, automatically incorporating transitions that occur on the nanosecond timescale of simulations, such as dynamic disorder of protons, or near-free rotations of groups of atoms, as well as -- if combined with constant-pressure simulations -- of thermal expansion of the lattice. 
In cases where GIPAW-DFT is known or suspected to provide an inadequate description of electronic structure, our integrated modeling workflow naturally extends to predictions of chemical shieldings (as well as energies and forces) based on quantum-chemical methods such as MP2~\cite{cervinka_ab_2018}, RPA~\cite{klimes_lattice_2016}, coupled cluster~\cite{gruber_applying_2018} or quantum Monte Carlo~\cite{zen_fast_2018} -- possibly using a baselined model and active-learning strategies that incorporate  uncertainty estimation for ML predictions~\cite{musil_2019_uncertainty, graselli2021}.

The framework we demonstrate promises to enable NMR-based crystal structure determination for flexible and complex (polymorphic) systems, and for compounds such as theophylline~\cite{engel2019bnmr} that contain a small number of independent $^1$H sites, which have this far eluded structure determination based on a direct GIPAW-DFT based approach. As the family of ML models we introduce here become transferable across different classes of compounds, we envisage the use of ML models to compute ensemble-averaged shifts to become common practice in NMR crystallography.

\section*{Supporting Information} \label{sec:si}

Additional details regarding the construction of the shielding models and sampling of nuclear fluctuations, and complete sets of shielding predictions for all polymorphs.
\section*{Acknowledgements} \label{sec:acknowledgements}

The authors would like to thank Lyndon Emsley for insightful comments on an early version of the manuscript, and Andrea Anelli for help setting up training of the ShiftML model.
VK acknowledges funding from the Swiss National Science Foundation (SNSF), Project $\text{P2ELP2}\_\text{191678}$. MC, VK and EAE acknowledge support from the NCCR MARVEL, funded by the SNSF.
EAE acknowledges funding from Trinity College, Cambridge.
EAE and VK acknowledges allocation of computing resources by CSCS under Project IDs s960 and s1000.

\bibliographystyle{ieeetr}

\begin{thebibliography}{10}

\bibitem{reif_2021}
B.~Reif, S.~E. Ashbrook, L.~Emsley, and M.~Hong, ``Solid-state nmr
  spectroscopy,'' {\em Nature Reviews: Methods Primers}, vol.~1, p.~2, 2021.

\bibitem{florian2013}
P.~Florian and D.~Massiot, ``{Beyond periodicity: Probing disorder in
  crystalline materials by solid-state nuclear magnetic resonance
  spectroscopy},'' {\em CrystEngComm}, vol.~15, p.~8623, 2013.

\bibitem{gee2000}
C.~H. Gee and W.~T. Raynes, ``{Nuclear motion effects on the 13C, 19F and 1H
  shielding in methyl fluoride},'' {\em Chem. Phys. Lett.}, vol.~330, p.~595,
  2000.

\bibitem{boehm2000}
M.~Böhm, J.~Schulte, and R.~Ram\'{i}rez, ``{Nuclear quantum effects in
  calculated NMR shieldings of ethylene; a Feynman path integral-ab initio
  study},'' {\em Chem. Phys. Lett.}, vol.~332, p.~117, 2000.

\bibitem{schulte2001}
J.~Schulte, R.~Ram\'{i}rez, and M.~C. Böhm, ``{Nuclear quantum effects in
  calculated NMR shieldings of benzene; a Feynman path integral study},'' {\em
  Mol. Phys.}, vol.~99, p.~1155, 2001.

\bibitem{ruden2003}
T.~A. Ruden, O.~B. Lutn{\ae}s, T.~Helgaker, , and K.~Ruud, ``{Vibrational
  corrections to indirect nuclear spin-spin coupling constants calculated by
  density-functional theory},'' {\em J. Chem. Phys.}, vol.~118, p.~9572, 2003.

\bibitem{dracinsky2009}
M.~Dra\u{c}ínsk\'{y}, J.~Kaminsk{\'y}, and P.~Bou\u{r}, ``{Relative importance
  of first and second derivatives of nuclear magnetic resonance chemical shifts
  and spin-spin coupling constants for vibrational averaging},'' {\em J. Chem.
  Phys.}, vol.~130, p.~094106, 2009.

\bibitem{zhou-wang20jcp}
S.~Zhou and L.~Wang, ``Quantum effects and {\textsuperscript{1}} {{H NMR}}
  chemical shifts of a bifurcated short hydrogen bond,'' {\em J. Chem. Phys.},
  vol.~153, p.~114301, Sept. 2020.

\bibitem{hohenberg1964}
.~P. Hohenberg and W.~Kohn, ``{Inhomogeneous electron gas},'' {\em Phys. Rev.},
  vol.~136, p.~B864, 1964.

\bibitem{kohn1965}
W.~Kohn and L.~J. Sham, ``{Self-consistent equations including exchange and
  correlation effects},'' {\em Phys. Rev.}, vol.~140, p.~A1133, 1965.

\bibitem{payne1992}
M.~C. Payne, M.~P. Teter, D.~C. Allan, T.~A. Arias, and J.~D. Joannopoulos,
  ``{Iterative minimization techniques for ab initio total-energy calculations:
  Molecular dynamics and conjugate gradients},'' {\em Rev. Mod. Phys.},
  vol.~64, p.~1045, 1992.

\bibitem{thon+09jcp}
T.~Thonhauser, D.~Ceresoli, A.~A. Mostofi, N.~Marzari, R.~Resta, and
  D.~Vanderbilt, ``A converse approach to the calculation of {{NMR}} shielding
  tensors,'' {\em J. Chem. Phys.}, vol.~131, no.~10, p.~101101, 2009.

\bibitem{bonhomme_2012}
C.~Bonhomme, C.~Gervais, F.~Babonneau, C.~Coelho, F.~Pourpoint, T.~Azais, S.~E.
  Ashbrook, J.~M. Griffin, J.~R. Yates, F.~Mauri, and C.~J. Pickard,
  ``First-principles calculation of nmr parameters using the gauge including
  projector augmented wave method: A chemist's point of view,'' {\em Chemical
  Reviews}, vol.~112, pp.~5733--5779, 2012.

\bibitem{pickard_2001_gipaw}
C.~J. Pickard and F.~Mauri, ``{All-electron magnetic response with
  pseudopotentials: NMR chemical shifts},'' {\em Physical Review B}, vol.~63,
  p.~245101, 2001.

\bibitem{yates_2007_gipaw}
J.~R. Yates, C.~J. Pickard, and F.~Mauri, ``{Calculation of NMR chemical shifts
  for extended systems using ultrasoft pseudopotentials},'' {\em Physical
  Review B}, vol.~76, p.~024401, 2007.

\bibitem{helgaker_2007}
T.~Helgaker and M.~Jaszu\'{n}ski, ``Density-functional and coupled-cluster
  singles-and-doubles calculations of the nuclear shielding and indirect
  nuclear spin-spin coupling constants of o-benzyne,'' {\em J. Chem. Theory
  Comput.}, vol.~3, p.~86, 2007.

\bibitem{hartman_2017}
J.~D. Hartman, A.~Balaji, and G.~J.~O. Beran, ``Improved electrostatic
  embedding for fragment-based chemical shift calculations in molecular
  crystals,'' {\em J. Chem. Theory Comput.}, vol.~13, p.~6043, 2017.

\bibitem{hartman_2018}
J.~D. Hartman and G.~J.~O. Beran, ``Accurate 13-c and 15-n molecular crystal
  chemical shielding tensors from fragment-based electronic structure theory,''
  {\em Solid State Nuclear Magnetic Resonance}, vol.~96, p.~10, 2018.

\bibitem{dittmer_2020}
A.~Dittmer, G.~L. Stoychev, D.~Maganas, A.~A. Auer, and F.~Neese, ``Computation
  of nmr shielding constants for solids using an embedded cluster approach with
  dft, double-hybrid dft, and mp2,'' {\em J. Chem. Theory Comput.}, vol.~16,
  p.~6950, 2020.

\bibitem{hofstetter_2017}
A.~Hofstetter and L.~Emsley, ``{Positional Variance in NMR Crystallography},''
  {\em J. Am. Chem. Soc}, vol.~139, p.~2573, 2017.

\bibitem{rossano2005}
S.~Rossano, F.~Mauri, C.~J. Pickard, and I.~Farnan, ``{First-principles
  calculation of 17O and 25Mg NMR shieldings in MgO at finite temperature:
  Rovibrational effect in solids},'' {\em J. Phys. Chem. B}, vol.~109, p.~7245,
  2005.

\bibitem{dumez2009}
J.-N. Dumez and C.~J. Pickard, ``{Calculation of NMR chemical shifts in organic
  solids: Accounting for motional effects},'' {\em J. Chem. Phys.}, vol.~130,
  p.~104701, 2009.

\bibitem{schmidt2005}
J.~Schmidt and D.~Sebastiani, ``{Anomalous temperature dependence of nuclear
  quadrupole interactions in strongly hydrogen-bonded systems from first
  principles},'' {\em J. Chem. Phys.}, vol.~123, p.~074501, 2005.

\bibitem{lee2007}
Y.~J. Lee, B.~Bing{\"o}l, T.~Murakhtina, D.~Sebastiani, W.~H. Meyer, G.~Wegner,
  and H.~W. Spiess, ``{High-resolution solid-state NMR studies of poly(vinyl
  phosphonic acid) proton-conducting polymer: Molecular structure and proton
  dynamics},'' {\em J. Phys. Chem. B}, vol.~111, p.~9711, 2007.

\bibitem{robinson2010}
M.~Robinson and P.~D. Haynes, ``{Dynamical effects in ab initio NMR
  calculations: Classical force fields fitted to quantum forces},'' {\em J.
  Chem. Phys.}, vol.~133, p.~084109, 2010.

\bibitem{gortari2010}
I.~D. Gortari, G.~Portella, X.~Salvatella, V.~S. Bajaj, P.~C.~A. {van der Wel},
  J.~R. Yates, M.~D. Segall, C.~J. Pickard, M.~C. Payne, and M.~Vendruscolo,
  ``{Time averaging of NMR chemical shifts in the MLF peptide in the solid
  state},'' {\em J. Am. Chem. Soc.}, vol.~132, p.~5993, 2010.

\bibitem{hass+12jacs}
A.~A. Hassanali, J.~Cuny, M.~Ceriotti, C.~J. Pickard, and M.~Parrinello, ``The
  fuzzy quantum proton in the hydrogen chloride hydrates,'' {\em J. Am. Chem.
  Soc.}, vol.~134, no.~20, pp.~8557--8569, 2012.

\bibitem{ceri+13pnas}
M.~Ceriotti, J.~Cuny, M.~Parrinello, and D.~E. Manolopoulos, ``Nuclear quantum
  effects and hydrogen bond fluctuations in water,'' {\em Proc. Natl. Acad.
  Sci. U. S. A.}, vol.~110, pp.~15591--15596, Sept. 2013.

\bibitem{dracinsky2013}
M.~Dra\u{c}ínsk\'{y} and P.~Hodgkinson, ``{A molecular dynamics study of the
  effects of fast molecular motions on solid-state NMR parameters},'' {\em
  CrystEngComm}, vol.~15, p.~8705, 2013.

\bibitem{dracinsky2014}
M.~Dra\u{c}ínsk\'{y} and P.~Hodgkinson, ``{Effects of quantum nuclear
  delocalisation on NMR parameters from path integral molecular dynamics},''
  {\em Chem. -- Eur. J.}, vol.~20, p.~2201, 2014.

\bibitem{dracinsky2012}
M.~Dra\u{c}ínsk\'{y} and P.~Bou\u{r}, ``{Vibrational averaging of the chemical
  shift in crystalline {$\alpha$}-glycine},'' {\em J. Comput. Chem.}, vol.~33,
  p.~1080, 2012.

\bibitem{monserrat2014}
B.~Monserrat, R.~J. Needs, and C.~J. Pickard, ``{Temperature effects in
  first-principles solid state calculations of the chemical shielding tensor
  made simple},'' {\em J. Chem. Phys.}, vol.~141, p.~134113, 2014.

\bibitem{reil-tkat14prl}
A.~M. Reilly and A.~Tkatchenko, ``Role of dispersion interactions in the
  polymorphism and entropic stabilization of the aspirin crystal,'' {\em Phys.
  Rev. Lett.}, vol.~113, p.~055701, July 2014.

\bibitem{rossi2016}
M.~Rossi, P.~Gasparotto, and M.~Ceriotti, ``{Anharmonic and Quantum
  Fluctuations in Molecular Crystals: A First-Principles Study of the Stability
  of Paracetamol},'' {\em Phys. Rev. Lett.}, vol.~117, p.~115702, 2016.

\bibitem{ko+18prm}
H.-Y. Ko, R.~A. DiStasio, B.~Santra, and R.~Car, ``Thermal expansion in
  dispersion-bound molecular crystals,'' {\em Phys. Rev. Materials}, vol.~2,
  p.~055603, May 2018.

\bibitem{kapil2019}
V.~Kapil, E.~A. Engel, M.~Rossi, and M.~Ceriotti, ``{Assessment of Approximate
  Methods for Anharmonic Free Energies},'' {\em J. Chem. Theory Comput.},
  vol.~15, p.~5845, 2019.

\bibitem{raim+19prm}
N.~Raimbault, V.~Athavale, and M.~Rossi, ``Anharmonic effects in the
  low-frequency vibrational modes of aspirin and paracetamol crystals,'' {\em
  Phys. Rev. Materials}, vol.~3, p.~053605, May 2019.

\bibitem{mark-ceri18nrc}
T.~E. Markland and M.~Ceriotti, ``Nuclear quantum effects enter the
  mainstream,'' {\em Nat. Rev. Chem.}, vol.~2, p.~0109, Feb. 2018.

\bibitem{katrusiak2010}
A.~Katrusiak, M.~Podsiadlo, and A.~Budzianowski, ``{Association CH$\cdots\pi$
  and No van der Waals Contacts at the Lowest Limits of Crystalline Benzene I
  and II Stability Regions},'' {\em Crystal Growth and Design}, vol.~10,
  p.~3461, 2010.

\bibitem{budzianowski2006}
A.~Budzianowski and A.~Katrusiak, ``{Pressure-frozen benzene I revisited},''
  {\em Acta Crystallographica}, vol.~B62, p.~94, 2006.

\bibitem{dodd1998}
I.~M. Dodd, S.~J. Maginn, M.~M. Harding, and R.~J. Davey, 1998.
\newblock {CSD Communication}.

\bibitem{leviel1981}
J.-L. Leviel, G.~Auvert, and J.-M. Savariault, ``{Succinic acid (neutron study,
  at 77$\deg$K) C$_4$H$_8$0$_4$},'' {\em Acta Crystallographica}, vol.~B37,
  p.~2185, 1981.

\bibitem{dawson2005}
A.~Dawson, D.~R. Allan, S.~A. Belmonte, S.~J. Clark, W.~I.~F. David, P.~A.
  McGregor, S.~Parsons, C.~R. Pulham, and L.~Sawyer, ``{Effect of High Pressure
  on the Crystal Structures of Polymorphs of Glycine},'' {\em Crystal Growth
  and Design}, vol.~5, p.~1415, 2005.

\bibitem{kapil_2018_ipi}
V.~Kapil, M.~Rossi, O.~Marsalek, R.~Petraglia, Y.~Litman, T.~Spura, B.~Cheng,
  A.~Cuzzocrea, R.~H. Mei{\ss}ner, D.~M.Wilkins, B.~A. Helfrecht, P.~Juda,
  S.~P. Bienvenue, W.~Fang, J.~Kessler, I.~Poltavsky, S.~Vandenbrande,
  J.~Wieme, and M.~Ceriotti, ``{i-PI 2.0: A universal force engine for advanced
  molecular simulations},'' {\em Computer Physics Communications}, vol.~236,
  p.~214, 2018.

\bibitem{behl-parr07prl}
J.~Behler and M.~Parrinello, ``Generalized {{Neural}}-{{Network
  Representation}} of {{High}}-{{Dimensional Potential}}-{{Energy Surfaces}},''
  {\em Phys. Rev. Lett.}, vol.~98, p.~146401, Apr. 2007.

\bibitem{bart+10prl}
A.~P. Bart{\'o}k, M.~C. Payne, R.~Kondor, and G.~Cs{\'a}nyi, ``Gaussian
  {{Approximation Potentials}}: {{The Accuracy}} of {{Quantum Mechanics}},
  without the {{Electrons}},'' {\em Phys. Rev. Lett.}, vol.~104, p.~136403,
  Apr. 2010.

\bibitem{chen+19pnas}
B.~Cheng, E.~A. Engel, J.~Behler, C.~Dellago, and M.~Ceriotti, ``Ab initio
  thermodynamics of liquid and solid water,'' {\em Proc. Natl. Acad. Sci. U. S.
  A.}, vol.~116, pp.~1110--1115, Jan. 2019.

\bibitem{jinn+19prl}
R.~Jinnouchi, J.~Lahnsteiner, F.~Karsai, G.~Kresse, and M.~Bokdam, ``Phase
  {{Transitions}} of {{Hybrid Perovskites Simulated}} by {{Machine}}-{{Learning
  Force Fields Trained}} on the {{Fly}} with {{Bayesian Inference}},'' {\em
  Phys. Rev. Lett.}, vol.~122, p.~225701, June 2019.

\bibitem{deri+20nc}
V.~L. Deringer, M.~A. Caro, and G.~Cs{\'a}nyi, ``A general-purpose
  machine-learning force field for bulk and nanostructured phosphorus,'' {\em
  Nat Commun}, vol.~11, p.~5461, Dec. 2020.

\bibitem{niu+20nc}
H.~Niu, L.~Bonati, P.~M. Piaggi, and M.~Parrinello, ``Ab initio phase diagram
  and nucleation of gallium,'' {\em Nat Commun}, vol.~11, p.~2654, Dec. 2020.

\bibitem{imbalzano2021gaas}
G.~Imbalzano and M.~Ceriotti, ``{Modeling the Ga/As binary system across
  temperatures and compositions from first principles},'' {\em Physical Review
  Materials}, vol.~5, p.~063804, 2021.

\bibitem{kapil2021}
V.~Kapil and E.~A. Engel, ``{A complete description of thermodynamic
  stabilities of molecular crystals},'' {\em arXiv e-prints},
  p.~arXiv:2102.13598, 2021.

\bibitem{perdew1996pbe0}
J.~P. Perdew, M.~Ernzerhof, and K.~Burke, ``{Rationale for mixing exact
  exchange with density functional approximations},'' {\em Journal of Chemical
  Physics}, vol.~105, p.~9982, 1996.

\bibitem{adamo1999pbe0}
C.~Adamo and V.~Barone, ``{Toward reliable density functional methods without
  adjustable parameters: The PBE0 model},'' {\em Journal of Chemical Physics},
  vol.~110, p.~6158, 1999.

\bibitem{tkatchenko2012mbd}
A.~Tkatchenko, R.~A. {Di Stasio}, R.~Car, and M.~Scheffler, ``{Accurate and
  efficient method for many-body van der waals interactions},'' {\em Physical
  Review Letters}, vol.~108, p.~236402, 2012.

\bibitem{ambrosetti2014mbd}
A.~Ambrosetti, A.~M. Reilly, R.~A. {Di Stasio}, and A.~Tkatchenko,
  ``{Long-range correlation energy calculated from coupled atomic response
  functions},'' {\em Journal of Chemical Physics}, vol.~140, p.~018A508, 2014.

\bibitem{matcloud21b}
V.~Kapil and E.~A. Engel, ``Dataset: {{Semi}}-local and hybrid functional
  {{DFT}} data for thermalised snapshots of polymorphs of benzene, succinic
  acid, and glycine.''
  \url{https://archive.materialscloud.org/record/2020.0026}, 2021.
\newblock (accessed 2021-01-07).

\bibitem{paruzzo_2018_shiftml}
F.~M. Paruzzo, A.~Hofstetter, F.~Musil, S.~De, M.~Ceriotti, and L.~Emsley,
  ``{Chemical shifts in molecular solids by machine learning},'' {\em Nature
  Communications}, vol.~9, p.~4501, 2018.

\bibitem{engel2019bnmr}
E.~A. Engel, A.~Anelli, A.~Hofstetter, F.~Paruzzo, L.~Emsley, and M.~Ceriotti,
  ``{A Bayesian approach to NMR crystal structure determination},'' {\em PCCP},
  vol.~21, p.~23385, 2019.

\bibitem{salager_2010_nmr}
E.~Salager, G.~M. Day, R.~S. Stein, C.~J. Pickard, B.~Elena, and L.~Emsley,
  ``{Powder crystallography by combined crystal structure prediction and
  high-resolution 1H solid-state NMR spectroscopy},'' {\em J. Am. Chem. Soc.},
  vol.~132, p.~2564, 2010.

\bibitem{hartman_2016}
J.~D. Hartman, R.~A. Kudla, G.~M. Day, L.~J. Mueller, and G.~J.~O. Beran,
  ``{Benchmark fragment-based 1H, 13C, 15N and 17O chemical shift predictions
  in molecular crystals},'' {\em Phys. Chem. Chem. Phys.}, vol.~18, p.~21686,
  2016.

\bibitem{dracinsky_2019}
M.~Dracinsky, P.~Unzueta, and G.~J.~O. Beran, ``{Improving the accuracy of
  solid-state nuclear magnetic resonance chemical shift prediction with a
  simple molecular correction},'' {\em Physical Chemistry Chemical Physics},
  vol.~21, p.~14992, 2019.

\bibitem{musil_2019_uncertainty}
F.~Musil, M.~J. Willatt, M.~A. Langovoy, and M.~Ceriotti, ``{Fast and Accurate
  Uncertainty Estimation in Chemical Machine Learning},'' {\em Journal of
  Chemical Theory and Computation}, vol.~15, p.~906, 2019.

\bibitem{jagannathan1987succinic}
N.~R. Jagannathan and C.~N.~R. Rao, ``{A $^{13}$C NMR spectroscopic study of
  the phase transitions of alkane dicarboxylic acids in the solid state},''
  {\em Chemical Physics Letters}, vol.~140, p.~46, 1987.

\bibitem{yamada2008glycine}
K.~Yamada, T.~Shimizu, T.~Yamazaki, and A.~Sato, ``{A Solid-state $^{17}$O NMR
  Study of $\beta$-Glycine: High Sensitivity of $^{17}$O NMR Parameters to
  Hydrogen-bonding Interactions},'' {\em Chemistry Letters}, vol.~37, p.~472,
  2008.

\bibitem{folliet2013glycine}
N.~Folliet {\em et~al.}, ``{Adsorption of Glycine in Mesoporous Silica through
  NMR Experiments Combined with DFT‐D Calculations},'' {\em Journal of
  Physical Chemistry C}, vol.~117, p.~4104, 2013.

\bibitem{cerreia2018glycine}
P.~Cerreia-Vioglio, G.~Mollica, M.~J. and. Colan E.~Hughes, P.~A. Williams,
  F.~Ziarelli, S.~Viel, P.~Thureau, and K.~D.~M. Harris, ``{Insights into the
  Crystallization and Structural Evolution of Glycine Dihydrate by In-Situ
  Solid‐State NMR Spectroscopy},'' {\em Angewandte Chemie: International
  Edition}, vol.~57, p.~6619, 2018.

\bibitem{paruzzo2019glycine}
F.~M. Paruzzo and L.~Emsley, ``{High-resolution $^{1}$H NMR of powdered solids
  by homonuclear dipolar decoupling},'' {\em Journal of Magnetic Resonance},
  vol.~309, p.~106598, 2019.

\bibitem{castiglia2019glycine}
F.~Castiglia, F.~Zevolini, G.~Riolo, J.~Brunetti, A.~{De Lazzari}, A.~Moretto,
  G.~Manetto, M.~Fragai, J.~Algotsson, J.~Even\"{a}s, L.~Bracci, A.~Pini, and
  C.~Falciani, ``{NMR Study of the Secondary Structure and Biopharmaceutical
  Formulation of an Active Branched Antimicrobial Peptide},'' {\em Molecules},
  vol.~24, p.~4290, 2019.

\bibitem{repository2}
E.~A. Engel, V.~Kapil, and M.~Ceriotti, ``{The importance of nuclear quantum
  effects for NMR crystallography},'' {\em Materials Cloud Archive},
  vol.~2021.118, 2021.

\bibitem{taylor2008}
R.~E. Taylor and C.~Dybowski, ``{Variable temperature NMR characterization of
  $\alpha$-glycine},'' {\em Journal of Molecular Structure}, vol.~889, p.~376,
  2008.

\bibitem{kapi+19jctc}
V.~Kapil, J.~Wieme, S.~Vandenbrande, A.~Lamaire, V.~Van~Speybroeck, and
  M.~Ceriotti, ``Modeling the {{Structural}} and {{Thermal Properties}} of
  {{Loaded Metal}}-{{Organic Frameworks}}. {{An Interplay}} of {{Quantum}} and
  {{Anharmonic Fluctuations}},'' {\em J. Chem. Theory Comput.}, vol.~15,
  pp.~3237--3249, May 2019.

\bibitem{cervinka_ab_2018}
C.~Červinka and G.~J.~O. Beran, ``Ab initio prediction of the polymorph phase
  diagram for crystalline methanol,'' {\em Chemical Science}, vol.~9,
  pp.~4622--4629, May 2018.

\bibitem{klimes_lattice_2016}
J.~Klimeš, ``Lattice energies of molecular solids from the random phase
  approximation with singles corrections,'' {\em The Journal of Chemical
  Physics}, vol.~145, p.~094506, Sept. 2016.

\bibitem{gruber_applying_2018}
T.~Gruber, K.~Liao, T.~Tsatsoulis, F.~Hummel, and A.~Grüneis, ``Applying the
  {Coupled}-{Cluster} {Ansatz} to {Solids} and {Surfaces} in the
  {Thermodynamic} {Limit},'' {\em Physical Review X}, vol.~8, p.~021043, May
  2018.

\bibitem{zen_fast_2018}
A.~Zen, J.~G. Brandenburg, J.~Klimeš, A.~Tkatchenko, D.~Alfè, and
  A.~Michaelides, ``Fast and accurate quantum {Monte} {Carlo} for molecular
  crystals,'' {\em Proceedings of the National Academy of Sciences}, vol.~115,
  pp.~1724--1729, Feb. 2018.

\bibitem{graselli2021}
G.~Imbalzano, Y.~Zhuang, V.~Kapil, K.~Rossi, E.~A. Engel, F.~Grasselli, and
  M.~Ceriotti, ``{Uncertainty estimation for molecular dynamics and
  sampling},'' {\em Journal of Chemical Physics}, vol.~154, p.~074102, 2021.

\end{thebibliography}

\end{document}